# Solar-type dynamo behaviour in fully convective stars without a tachocline

Nicholas J. Wright[1] & Jeremy J. Drake[2]

**In solar-type stars (with radiative cores and convective envelopes), the magnetic field powers star spots, flares and other solar phenomena, as well as chromospheric and coronal emission at ultraviolet to X-ray wavelengths. The dynamo responsible for generating the field depends on the shearing of internal magnetic fields by differential rotation[1,2]. The shearing has long been thought to take place in a boundary layer known as the tachocline between the radiative core and the convective envelope[3]. Fully convective stars do not have a tachocline and their dynamo mechanism is expected to be very different[4], although its exact form and physical dependencies are not known. Here we report observations of four fully convective stars whose X-ray emission correlates with their rotation periods in the same way as in Sun-like stars. As the X-ray activity – rotation relationship is a well-established proxy for the behaviour of the magnetic dynamo, these results imply that fully convective stars also operate a solar-type dynamo. The lack of a tachocline in fully convective stars therefore suggests that this is not a critical ingredient in the solar dynamo and supports models in which the dynamo originates throughout the convection zone.**

Stars across the Hertzsprung–Russell diagram are known to emit X-rays, with only a few exceptions. In main-sequence solar-type and low-mass stars, the X-rays arise from a magnetically confined plasma known as a corona that reaches temperatures of several million kelvin[5]. Coronal X-ray emission is ultimately powered by the dissipation of the magnetic fields generated by an interior magnetic dynamo. A close relation between the surface magnetic flux and X-ray radiance based on both solar and stellar observations that span several decades for both quantities[6] indicates that X-ray emission is a reliable proxy for magnetic activity. The dynamo is thought to be driven in part by differential rotation in the interior of the star[1,2], which itself is generated by the action of the Coriolis force on the rotating convective envelope, but the detailed mechanism remains to be properly understood[7]. The relationship between stellar rotation and tracers of magnetic activity is therefore an important probe of the stellar dynamo.

In Sun-like stars X-ray emission is observed to increase monoton- ically with increasing stellar rotational velocity for periods exceeding a few days[8,9]. This relationship is often quantified in terms of the dependence of the ratio of stellar luminosity expended in X-rays to bolometric luminosity $L_X/L_{bol}$, on the Rossby number, which is defined as $Ro = P_{rot}/\tau$, the ratio of the stellar rotation period and the mass- dependent convective turnover time[10]. A recent study[9] using the largest available sample of 824 solar- and late-type stars fitted the rotation– activity relation as $L_X/L_{bol} = 5.3 \times 10^{-6}\, Ro^{-2.7}$.

This relationship has been observed in stars from late F-type through to early M-type, that is, those with radiative cores and convective envelopes. The interface layer between these two regions, named the tachocline, is believed to play an important role in the generation of the magnetic field[3]. Shear between the rigidly rotating core and the differential rotation of the convective envelope with latitude is thought to amplify and store the magnetic field[11], generating what is known as an α–Ω dynamo, named for the interplay between cyclonic eddies (the α effect) and the shearing of the field (the Ω effect).

Others have argued that the latitudinal and radial gradients of the angular velocity in the convection zone may be sufficient for global dynamo action[12,13]. In the Sun, the extremely strong levels of radial shear just beneath the surface are actually higher than in the tachocline, making it plausible that the solar dynamo is distributed across the convection zone rather than confined to the tachocline[12].

Despite this hypothesis it has been widely accepted that magnetic structures in the convection zone would be disrupted by magnetic buoyancy or turbulent pumping, preventing large-scale magnetic fields from being established there[14]. However, some recent three- dimensional magneto-hydrodynamic simulations without a tachocline have produced persistent magnetic wreaths in the convection zone[15] and shown that it is possible to produce large-scale magnetic fields in stellar convection layers[16,17].

For very fast rotators, the rotation–activity relationship has been found to break down, with X-ray luminosity reaching a saturation level of approximately $L_X/L_{bol} \approx 10^{-3}$, independent of the spectral type[18]. This saturation

---

[1] Astrophysics Group, Keele University, Keele, ST5 5BG, UK
[2] Harvard-Smithsonian Center for Astrophysics, 60 Garden Street, Cambridge, MA 02138, USA

level is reached at a Rossby number of approximately 0.13 ± 0.02 (ref. 9), corresponding to a rotation period that increases towards later spectral types, from 2 days for a star similar to the Sun to up to about 20 days for low-mass M dwarfs, and is also seen in both chromospheric emission and magnetic field measurements. It is unclear whether this is caused by a saturation of either the dynamo mechanism or the transport of the magnetic flux to the corona, a change in the type of dynamo at work within the star[9] or because coronal X-ray emission itself becomes insensitive to the strength of the magnetic field, the energy of which is then dissipated in other ways.

Main-sequence stars later than spectral type M3–3.5 (M < 0.4M$_\odot$, where M is the mass of the star and M$_\odot$ is the mass of the Sun) are predicted to be fully convective and therefore do not possess a tachocline. If the tachocline is critical to the operation of a solar-type dynamo, fully convective stars should not be able to sustain such a dynamo. Instead, it is generally thought that they generate magnetic fields entirely by helical turbulence[4]. Nevertheless, observations indicate that stars throughout the M-type spectral range exhibit high magnetic field strengths[19] and high fractional X-ray luminosities[9]. In fact, no discernible difference in magnetic activity properties has been identified on either side of the fully convective boundary.

One problem with existing studies is that nearly all fully convective stars that have so far been studied have saturated levels of X-ray emission[9]. It is therefore unclear whether these stars all exhibit saturated levels of X-ray emission (possibly hinting at some facet of their dynamo mechanism), or whether slower rotators follow a more solar-like rotation–activity relationship. Studies of magnetic activity in slowly rotating fully convective stars have so far been lacking, adding to the uncertainty about their dynamo mechanism. Despite this, fully convective stars are common and, given their spin-down times of a few billion years[20], at least half of all such stars are expected to be slow rotators.

Understanding the dynamo mechanism in these slowly rotating stars is important for various astrophysical problems, including the dynamo-driven angular momentum loss rate of low-mass stars, the particle and photon radiation environment of exoplanets and the notorious period gap in the cataclysmic variables. The lack of cataclysmic variables with periods in the 2–3 h range is often attributed to a change in rotational spin-down for fully convective stars with diminished magnetic dynamos and stellar winds[21]. Recent numerical simulations of stellar winds suggest that increasing the complexity of the surface magnetic morphology can also suppress angular momentum loss and spin-down without requiring any change in the total surface magnetic flux[22].

Four slowly rotating, fully convective M-type stars were observed by either NASA's Chandra X-ray Observatory or the ROSAT satellite. These stars are of spectral type M4–5.5, well beyond the fully convective boundary (M3–3.5) and are therefore genuine fully convective stars. Figure 1 illustrates the traditional rotation–activity diagram showing the fractional X-ray luminosity as a function of the Rossby number for all of the stars from the most recent large-scale study of the rotation–activity relationship in solar-type and low-mass stars[9]. The positions of the four slowly rotating fully convective stars studied in this work are also shown, and are in excellent agreement with the rotation–activity relationship of partly convective stars.

The observations presented here provide clear evidence for unsaturated X-ray emission in fully convective stars and quantification of their rotation–activity relationship. The results show that fully convective stars, at least when they have spun down sufficiently, operate a dynamo that exhibits a rotation–activity relationship that is indistinguishable from that of solar-type stars. As the dynamo action in fully convective stars is expected to be different from that in solar-type stars owing to their lack of a tachocline[4], where the magnetic field is thought to be amplified by radial shear, this is a surprising finding.

The most direct conclusion from these observations is that both partly and fully convective stars operate very similar rotation-dependent dynamos in which the tachocline is not a vital ingredient and differential rotation combined with the action of the Coriolis force is sufficient[7]. This implies that current models for the solar dynamo that rely on the tachocline layer to amplify the magnetic field are incorrect, and lends weight to recent three-dimensional magneto-hydrodynamic simulations without a tachocline that produce large-scale magnetic fields entirely within the convective layers[23,24]. Recent studies with mean-field dynamo models have also suggested that differential rotation in the convection zone may play a greater role in the generation of the magnetic field than does the tachocline[25,26].

Another alternative possibility is that fully convective stars are able to generate a purely turbulent dynamo that exhibits a rotation–activity relationship that is similar to that in partly convective stars, but by a different mechanism that does not rely on a shear layer. Existing dynamo simulations for fully convective stars succeed in generating magnetic fields, but are unable to predict their behaviour as a function of the rotation rate[17]. However, it seems unlikely that both partly and fully convective stars would have the same rotation–activity relationship (requiring both their dynamo efficiency and rotational dependence to behave in the same way) without their dynamo mechanisms sharing a major feature.

A third possibility is that convection in the cores of fully convective stars could be magnetically suppressed[27], leading to the existence of a solar-like tachocline, although some studies suggest that convection would not be completely halted, only made less efficient[28]. Furthermore, the field strengths that are necessary for such a transition are $10^7$–$10^8$ G (refs 28, 29), orders of magnitude larger than the fields thought to exist in the solar interior and at levels that simulations suggest are impossible to maintain[30].

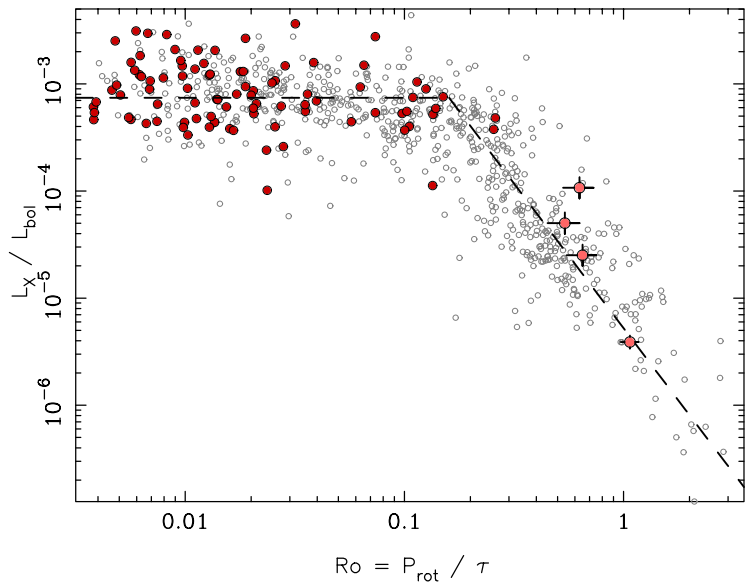

*Figure 1. Rotation–activity relationship diagram for partly and fully convective stars.* Fractional X-ray luminosity, $L_X/L_{bol}$, plotted against the Rossby number, $Ro = P_{rot}/\tau$, for 824 partly (grey circles) and fully (red circles) convective stars from the most recent large compilation of stars with measured rotation periods and X-ray luminosities[7]. The best-fitting saturated (horizontal) and unsaturated (diagonal) rotation–activity relationships from that study are shown as black dashed lines. The four slowly rotating fully convective M dwarfs studied here are shown in light red (error bars indicate 1 standard deviation). The uncertainties for the other data points are not quantified but will be comparable to the M dwarfs for the Rossby number and approximately twice as large for $L_X/L_{bol}$.

**Acknowledgments:** N.J.W. acknowledges a Royal Astronomical Society Research Fellowship and an STFC Ernest Rutherford Fellowship. J.J.D. was supported by NASA contract NAS8-03060 to the Chandra X-ray Center. The authors would like to thank Jonathan Irwin, Rob Jeffries and Andrew West for assistance and comments on an early draft of this paper. This research has made use of the Vizier and SIMBAD databases (operated at CDS, Strasbourg, France).

**Author Contributions:** N.J.W. reduced the Chandra observations, measured the X-ray fluxes and made the necessary calculations to plot the sources in Figure 1. N.J.W. and J.J.D. wrote the interpretation and discussion of the results.

**Author Information:** Reprints and permissions information is available at www.nature.com/reprints. The authors declare no competing financial interests. Correspondence and requests for materials should be addressed to nick.nwright@gmail.com.

**Methods**

**New observations.** Two targets were chosen from the MEarth Transit survey[20] of fully convective stars: G184-31 and GJ 3253, of spectral types M4.5[31] and M5[32] with rotation periods of 83.8 days (ref. 33) and 78.8 days (ref. 20), respectively.

The targets were observed with the Advanced CCD Imaging Spectrometer (ACIS)[34] on the Chandra X-ray Observatory[35] using the ACIS-S (spectroscopic) CCD array. Observations were performed in 'very faint' mode and were placed on the back-illuminated S3 (the 3rd CCD chip in the ACIS-S spectroscopic array) chip (owing to its higher sensitivity to soft X-rays. The observations were performed on 6 December 2012 and 25 September 2013 for G184-31 and GJ 3253 respectively, with exposure times of 8.0 ks and 21.0 ks.

Observations were processed using the CIAO 4.5 software tools[36] and the CALDB 4.5.8 calibration files following standard procedures. The two sources were clearly identified and detected with significances of $5.0\sigma$ and $10.2\sigma$ at their expected positions. Point-source extraction was performed using CIAO 4.5. The total number of net counts was measured to be $35^{+7}_{-6}$ and $124^{+12}_{-11}$ for G184-31 and GJ 3253, respectively. Light curves were constructed for both sources to search for high levels of variability that might inflate the quiescent flux level measured, but no significant variability was detected.

Thermal plasma X-ray spectral models were fitted to the extracted spectra using XSPEC[37] version 12.6.0 and compared to APEC[38] single-temperature optically thin model spectra of an absorbed thermal plasma in collisional ionization equilibrium, allowing the plasma temperature ($k_BT$, where $k_B$ is Boltzmann's constant and T is temperature) and the hydrogen column density ($N_H$) to vary freely. A grid of initial thermal plasma temperatures covering the range $k_BT = 0.1–3.0$ keV was used to prevent fitting to local minima. The model with the lowest Cash statistic[39] was selected as the best fit for each source (the Cash statistic is an application of the likelihood ratio test that is suitable for low-signal data). The best-fitting thermal plasma temperatures were found to be $k_BT = 0.78 \pm 0.13$ keV and $0.30 \pm 0.05$ keV for G184-31 and GJ 3253, respectively, consistent with the values found for other M-type dwarf stars. The fitted hydrogen column density is consistent with no absorption, as expected for the proximity of these sources.

Absorption-corrected broadband (0.5–8.0 keV) fluxes of $F_X = (2.08 \pm 0.38) \times 10^{-14}$ erg/s/cm$^2$ and $(4.35 \pm 0.41) \times 10^{-14}$ erg/s/cm$^2$ were calculated from the model fits for G184-31 and GJ 3253 respectively. Combined with their known parallax distances[31,32] the fluxes were used to calculate X-ray luminosities in the ROSAT band (0.1–2.4 keV), for consistency and ease of comparison with previous studies[7]. Fractional X-ray luminosities were then calculated from the observed J-band magnitudes and the appropriate bolometric corrections[40].

**Literature data.** We searched the literature for fully convective stars (spectral type M4 or later) with existing measured rotation periods and X-ray luminosities, excluding any with short rotation periods ($P_{rot} < 20$ days), which would place the object in the saturated regime of Fig. 1. Two stars were found that met our criteria: GJ 699 (Barnard's star, M4V; ref. 41) and GJ 551 (Proxima Centauri, M5.5V; ref. 42) with rotation periods[43] and X-ray luminosity values[44] existing in the literature. Rossby numbers and fractional X-ray luminosities for these stars were calculated as for the two newly observed stars.

We also uncovered a number of stars that are close to the convective boundary in the literature[45,46] (spectral types M3 or M3.5). The spectral types and colours[9] of these stars imply that they are partly convective and so they are shown in Fig. 1 as grey dots.

**Sample size.** No statistical methods were used to predetermine sample size.

**Code availability.** The CIAO code used to reduce the Chandra X-ray Observatory data are available at http://cxc.cfa.harvard.edu/ciao and the associated calibration database can be found at http://cxc.cfa.harvard.edu/caldb. The XSPEC code used to perform X-ray spectral fitting is available at https://heasarc.gsfc.nasa.gov/xanadu/ xspec.